\documentclass[12pt]{article}

\usepackage{sbc-template}
\usepackage[latin1]{inputenc} 
\usepackage{graphicx}
\usepackage{url}
\usepackage{natbib}
\usepackage{hyperref}

\newcommand{\mytitle}{The Performance of Paxos and Fast Paxos}

\hypersetup{
     pdftitle={\mytitle}, 
     pdfauthor={Gustavo M. D. Vieira, Luiz E. Buzato},
     pdfdisplaydoctitle=true,
     hidelinks
}

\bibpunct{[}{]}{,}{a}{}{,}
\renewcommand{\cite}[2][]{\citep[#1]{#2}}

\title{\mytitle}

\author{Gustavo M.  D.  Vieira\inst{1}\thanks{Supported by CNPq grant
    142638/2005-6.}  \and Luiz E.  Buzato\inst{1}\thanks{Supported by
    CNPq grant 201934/2007-8.}}

\address{Instituto de Computação, Unicamp \\
         Caixa Postal 6176 \\
         13083-970 Campinas, São Paulo, Brasil
         \email{\{gdvieira, buzato\}@ic.unicamp.br}}

\begin{document}

\maketitle  

\begin{abstract}
  Paxos  and Fast  Paxos  are optimal  consensus  algorithms that  are
  simple and elegant, while suitable for efficient implementation.  In
  this  paper,  we  compare  the  performance of  both  algorithms  in
  failure-free  and  failure-prone  runs  using  Treplica,  a  general
  replication toolkit  that implements  these algorithms in  a modular
  and  efficient manner.  We have  found that  Paxos  outperforms Fast
  Paxos for small  number of replicas and that  collisions are not the
  cause of this performance difference.
\end{abstract}

\section{Introduction}

The   construction  of  highly   available  asynchronous   systems  is
intrinsically linked to solutions to the problem of consensus, because
this problem is equivalent to a very powerful communication primitive:
total   order   broadcast~\cite{chandra96}.    Among   the   consensus
algorithms     available,      Paxos~\cite{lamport98}     and     Fast
Paxos~\cite{lamport06a} have recently been used to implement important
systems~\cite{chandra07} for at least  the following reasons: (i) they
implement  uniform consensus; (ii)  they are  simple and  elegant; and
(iii)  they are  efficient.  In  theory, the  number  of communication
rounds and the message complexity  required by Paxos and Fast Paxos to
reach consensus  should be the  determinant factors of  their expected
performance~\cite{deprisco00}.   Fast Paxos, with  smaller theoretical
latency,  should be  faster and  Paxos  should be  more resilient,  by
tolerating a larger number of failures.  Fast Paxos reduces latency by
being  optimistic,  that  is,  if  the  messages  exchanged  to  reach
consensus happen to be in a favorable order, then it is fast.  This is
the picture painted by theory.  Practice can paint different pictures.
Junqueira et  al.~\cite{junqueira07} have pinpointed  a scenario where
Paxos  shows  a smaller  overall  consensus  latency,  if one  of  the
communication  steps is  always much  slower than  the  others.  Their
results  serve  well  to  illustrate that  determining  the  practical
performance of  Fast Paxos and Paxos  can be a  challenging task whose
answers depend on careful experimentation.

In this  work, we  address the challenge  of assessing Paxos  and Fast
Paxos efficiencies in  practice in a LAN setting.  We decided to start
our  study  in the  LAN  environment because  it  houses  most of  the
applications   requiring  the   use  of   Paxos~\cite{chandra07}.  The
evaluation presented here was only possible because we have programmed
and tested both algorithms while building Treplica~\cite{vieira08a}, a
general replication  toolkit that can be instrumented  to generate the
indicators  necessary to  assess  the performance  of these  consensus
algorithms.  Our  assessment method  is based on  looking at  what the
theory  prescribes  for the  behaviour  of  the  algorithms to  design
experiments that are intended to observe whether or not the prescribed
behaviour occurs  in practice.   Examples of aspects  assessed include
number  of messages  ordered, latency  of messages,  quorum  sizes and
collisions.  We  have experimentally found, among  other results, that
Paxos    outperforms    Fast     Paxos    for    small    number    of
processes. Surprisingly, this isn't  caused by unjustified optimism in
Fast Paxos,  but by the  network and the  extra load generated  by the
uncoordinated activities of its processes.

The   remaining    of   the    paper   is   organized    as   follows.
Section~\ref{sec:theo}  details the theoretical  aspects of  Paxos and
Fast Paxos  and the key  differences between them.   The prescriptions
listed   here  were  used   as  a   guide  for   the  design   of  the
experiments. Section~\ref{sec:eval}  describes our experimental setup,
the  experiments, and  the  results obtained.   It  also contains  our
assessment of the results and  what they mean when contrasted with the
theoretical  predictions.  We  conclude the  paper with  a  section on
related work and a few concluding remarks.

\section{Theory}
\label{sec:theo}

Informally, the \emph{consensus} problem  consists in all processes in
a  distributed system  proposing an  initial value  and  all processes
eventually deciding on the same value from the ones proposed.  In this
section we  describe how Paxos and  Fast Paxos solve  consensus and we
argue  that there  are many  factors found  in real  systems  that can
affect these performance expectations.

\subsection{Paxos and Fast Paxos}

We give here a brief description  of Paxos and Fast Paxos, to create a
guide for  the experiments.  Full descriptions of  both algorithms can
be found  in~\cite{lamport06a}, including the computational and
failure models assumed by them.  Processes in the  system are reactive
agents  that can perform  multiple roles:  a \emph{proposer}  that 
proposes values, an  \emph{acceptor} that chooses a single  value, or a
\emph{learner} that learns what value has been chosen.

To solve  consensus, Paxos agents execute multiple  rounds, each round
has  a \emph{coordinator}  and is  uniquely identified  by  a positive
integer.  Proposers send their  proposed value to the coordinator that
tries to  reach consensus  on it  in a new  round. The  coordinator is
responsible for that round and is  able to decide, by applying a local
rule, if previous rounds were successful or not. The local rule of the
coordinator  is based  on quorums  of acceptors  and requires  that at
least $\lfloor N/2 \rfloor + 1$  processes take part in a round, where
$N$ is  the total number of processes  in the system~\cite{lamport06a,
  vieira08b}.  Each round progresses through two phases with two steps
each:
\begin{itemize}
\item In  Phase 1a  the coordinator sends  a message  requesting every
  acceptor  to participate  in  round $i$.   An  acceptor accepts  the
  invitation if it has not already accepted to participate in round $j
  \geq i$, otherwise it declines the invitation by simply ignoring it.
\item  In Phase  1b every  acceptor that  has accepted  the invitation
  answers  to the  coordinator with  a reply  that contains  the round
  number and the  value of the last  vote it has cast for  a value, or
  \textsl{null} if it has not voted. 
\item  In Phase  2a,  if the  coordinator  of round  $i$ has  received
  answers from a  quorum of acceptors then it  executes its local rule
  on the set of values suggested  by acceptors in Phase 1b and picks a
  single value $v$.  It then asks the acceptors to cast a vote for $v$
  in round $i$, if $v$ is not \textsl{null}, otherwise the coordinator
  is  free to  pick any  value  and picks  the value  proposed by  the
  proposer.
\item In Phase  2b, after receiving a request to cast  a vote from the
  coordinator, acceptors can either cast  a vote for $v$ in round $i$,
  if they  have not  voted in  any round $j  \geq i$,  otherwise, they
  ignore the  vote request.  Votes  are cast by sending  them together
  with the round identifier to the learners.
\item Finally, a  learner learns that a value $v$  has been chosen if,
  for some round  $i$, it receives Phase 2b messages  from a quorum of
  acceptors announcing that they have all voted for $v$ in round $i$.
\end{itemize}

Fast Paxos  changes Paxos by  allowing the proposers to  send proposed
values  directly  to  the  acceptors.   To achieve  this,  rounds  are
separated in  \emph{fast} rounds and \emph{classic}  rounds.  Fast and
classic rounds  have different quorums  with properties such  that the
local rule  of the coordinator is  still able to detect  if a previous
round was successful.  These quorums  are larger than the ones used by
Paxos and can assume many  values that satisfy the requirements of the
local rule.  In  particular, it is possible to  minimize the number of
processes  in a  fast quorum  ensuring that  both a  fast  and classic
quorums contain  $\lfloor 2N/3 \rfloor + 1$  processes. Another option
is to  minimize the number  of processes in classic  quorums requiring
the same number  of processes as in Paxos ($\lfloor  N/2 \rfloor + 1$)
but   requiring   $\lceil  3N/4   \rceil$   processes   in  the   fast
quorums~\cite{lamport06a, vieira08b}.   A Fast Paxos  round progresses
similarly to a Paxos round, except that Phase 2 is changed:
\begin{itemize}
\item  In Phase  2a,  if the  coordinator  of round  $i$ has  received
  answers from a  quorum of acceptors then it  executes its local rule
  on the set of values suggested  by acceptors in Phase 1b and picks a
  single value $v$.  It then asks the acceptors to cast a vote for $v$
  in round  $i$, if $v$ is  not \textsl{null}, otherwise, if  $i$ is a
  fast  round  the  coordinator  sends  a \emph{any}  message  to  the
  proposers indicating that any value  can be chosen in round $i$.  In
  this case,  the proposers can ask  the acceptors directly  to cast a
  vote for a value $v$ of their choice in round $i$.
\item In Phase  2b, after receiving a request to cast  a vote from the
  coordinator (if the  round is classic) or from  one of the proposers
  (if the round is fast), acceptors  can either cast a vote for $v$ in
  round  $i$,  if  they have  not  voted  in  any  round $j  \geq  i$,
  otherwise, they ignore the vote request.
\end{itemize}

The  above description  of  both algorithms  considers  only a  single
instance of  consensus.  However,  these algorithms are  more commonly
used to deliver a set of totally ordered messages, where a sequence of
repeated instance  of consensus maps  to a predefined position  in the
message ordering.   In this case,  it is possible  to run Phase  1 and
Phase 2a only once for all still unused instances.  This factorization
of  phases  is  carried  out  immediately  after  the  election  of  a
coordinator.  At this point, most  of the consensus instances have not
been started yet, allowing the  coordinator in Paxos to ``save'' these
instances for future use or, in Fast Paxos, allowing it to send Phase 2a
\emph{any} messages.

The improvement  brought about by  this factorization allows  Paxos to
achieve consensus in three communication rounds and Fast Paxos in only
two  communication   rounds.   Moreover,   in  Fast  Paxos   once  the
coordinator sends  the \emph{any}  messages, consensus can  be reached
without the need  of further coordinator intervention.  Unfortunately,
Fast Paxos cannot always be fast.  Proposers can propose two different
values concurrently, in this case their proposals may collide.  Also,
process and communication failures  may block a round from succeeding.
Different  recovery  mechanisms  can   be  implemented  to  deal  with
collisions and  failures, but eventually  the coordinator intervention
may be  necessary to start a new  classic round~\cite{lamport06a}.  In
both algorithms, any process can act  as the coordinator as long as it
follows the  rule for choosing  a value, if  any, that is  proposed in
Phase 2a.  The  choice of coordinator and the decision  to start a new
round of consensus are made relying in some timeout mechanism, as both
Paxos  and Fast  Paxos  assume a  partially synchronous  computational
model to ensure liveness.

\subsection{Performance Expectations}
\label{sec:expectations}

Before discussing  the performance  characteristics of Paxos  and Fast
Paxos experimentally,  it is useful  to map the theoretical  notion of
broadcast  onto the  actual  primitive available  in the  experimental
setup:  high speed wired  local area  networks (LAN).   The technology
most often  used to implement  these LANs is  Ethernet, in one  of its
several variations.  Because of  this heritage, it is commonly assumed
that LANs  use some  sort of shared  medium that must  be collectively
managed by the  stations connected to the network.   As a consequence,
LANs messages can  be broadcast to all stations  with the same latency
of  sending a  single message  and, due  to the  shared nature  of the
medium, only one of such broadcasts can happen at the same time.  This
characteristic is very  desirable, specially for optimistic algorithms
such  as Fast  Paxos.   However,  not all  variants  of Ethernet  work
through a  shared medium.  In  particular, 100Mbps and  1Gbps Ethernet
are usually implemented with a full-duplex dedicated twisted-pair link
connecting each  station to  a central switch  in a star  topology. In
these networks communication is centrally arbitrated by the switch and
there is  no need for stations  to manage access to  the medium.  This
setup has  many advantages to point to  point communication, including
full-duplex  communication  at full  speed  and  a maximum  aggregated
bandwidth larger  the individual bandwidth of any  link.  Broadcast is
still available,  but it is  not as straightforward  as it was  in the
shared  medium case.   In these  networks broadcast  is just  a single
message multiplied  by the switch and  put in the  dedicated medium of
each  station.  As  such,  every  one of  these  messages traverses  a
different queue and can  potentially be ordered differently from other
concurrent broadcasts  and unicasts. Moreover, it is not uncommon for
IP stacks  to  deliver  locally  a broadcast  message even   before it
reaches the network interface.

Within this  environment, what are the main  differences between Paxos
and  Fast Paxos  concerning  the \emph{expected}  performance of  both
algorithms?  Paxos  requires 3 communication rounds  for each instance
of   consensus   while  Fast   Paxos   needs   only  2   communication
rounds. Moreover, Fast Paxos  doesn't require the active participation
of   a  single  process,   the  coordinator,   in  all   instances  of
consensus. However, Fast Paxos  requires the participation of a larger
number of active processes than Paxos and the performance advantage of
Fast Paxos is  only realized in the optimistic case  where there is no
conflict.   Considering  these properties,  it  might  be tempting  to
conclude that as long as  the optimistic ordering of messages expected
by Fast Paxos holds this algorithm has the performance advantage.  For
each of  the potential advantages of  each algorithm we  list now some
reasons why this isn't necessary true:

\begin{description}
\item[Communication  rounds:]  The  main  claim  for  the  theoretical
  performance  of Fast  Paxos  is that  two  communication rounds  are
  better than  three.  However,  both Paxos and  Fast Paxos  contain a
  communication  step where  all  processes in  a  quorum broadcast  a
  message at the same time. No matter how efficient the switch is, all
  these broadcasts will have to  be serialized as they are transferred
  to all destination ports and they will be received as $k$ individual
  messages. In this  case, we can conceivably fold  in a communication
  round all  processing latency, but the  propagation and transmission
  latency  must  be  counted  individually.   That  is,  communication
  \emph{complexity} is important.
\item[Single coordinator:] All Paxos  messages must be relayed through
  a single coordinator. Although this  process isn't a single point of
  failure, it is a potential performance choke point. Fast Paxos might
  perform  better  if  load  on  the  coordinator  is  high,  but  the
  centralizing nature of the coordinator can act as more robust way to
  decide   on   an   order   for   the  messages   than   relying   on
  chance.
\item[Larger quorums:] Fast Paxos requires larger quorums and this has
  the  direct consequence  that the  algorithm tolerates  less process
  failures.  Depending on  the selection  of quorums  Fast  Paxos can
  revert to Paxos quorums ($\lfloor  N/2 \rfloor + 1$) if consensus is
  not optimistically  reached, but  this requires even  larger quorums
  for    the   optimistic   case.     This   fact    has   performance
  implications.   Larger   quorums  require   more   messages  to   be
  successfully  and  timely delivered  for  consensus  to be  reached,
  making  Fast  Paxos  vulnerable   to  network  overload  and  timing
  violations.
\item[Collisions:]  Fast  Paxos  is  optimistic. It  succeeds  in  two
  communication rounds as long as messages are naturally ordered. But,
  in switched LANs broadcasts are implemented as many messages send to
  each station, not necessarily ordered.   If only a majority of these
  messages  are ordered, consensus  will be  reached but  will require
  more  messages to  be timely  received. If  not even  a  majority of
  messages  is ordered,  a  collision occurred  and  consensus is  not
  possible. There is  nothing in the network that  orders messages. If
  they  arrive ordered  it  is more  likely  that they  were not  sent
  concurrently  in the first  place, thus  collisions increase  as the
  message rate increases~\cite{pedone03a}.
\end{description}

Observing the uncertainties related to each supposed advantage of Fast
Paxos,  it  is  possible  to  reach  the  conclusion  that  these  two
algorithms are basically incomparable without a clear characterization
of the  network properties. In  the next section  we present a  set of
experiments designed to extract  data on this characterization for our
target high speed local networks.

\section{Practice}
\label{sec:eval}

This  section presents the  basic organization  of Treplica  and where
Paxos and Fast Paxos where used in the toolkit.  Here, we also present
the experiments  we have carried out  to assess Fast  Paxos and Paxos,
their  results, and  what they  indicate in  relation to  the expected
behaviour indicated by theory.

\subsection{Treplica}

Treplica is  a replication toolkit that simplifies  the development of
high-available applications by  making transparent the complexities of
dealing with  replication and persistence.  We present  here the basic
organization of  Treplica and  where Paxos and  Fast Paxos fit  in the
toolkit.    Additional   information   on   Treplica  can   be   found
in~\cite{vieira08a}.   Treplica supports  the  construction of  highly
available  applications  through  either the  asynchronous  persistent
queue  or the  state machine  programming  interfaces.  A  queue is  a
\emph{totally   ordered}  collection   of  objects   with   the  usual
\emph{enqueue} and  \emph{dequeue} operations.  Persistence guarantees
that a process can crash, recover and bind again to its queue, certain
that the queue has preserved its  state and that it has not missed any
additional  enqueues made  by  any active  replicas.  An  asynchronous
persistent queue maintains  a history of the objects  it has ever held
since its creation.  Thus, by relying on the total order guaranteed by
the  queue and  in the  fact  that queues  are persistent,  individual
processes can  become active  replicas while remaining  stateless; the
persistence of their state has  been delegated to the queue. The state
machine  programming  interface  leverages  the persistent  queues  to
provide  a simple  abstraction of  an object  that only  changes state
through  deterministic  command  objects.   To use  this  abstraction,
applications     must     adhere      to     the     state     machine
approach~\cite{lamport78,schneider90}.

To provide  these two  programming abstractions and  still be  able to
provide reasonable  performance, Treplica uses  a \emph{uniform} total
order delivery mechanism built on  top of Paxos. The uniformity of the
consensus  component is  fundamental  to the  efficiency of  Treplica.
Usually,  uniform   consensus  algorithms  are   more  expensive  than
non-uniform  consensus algorithms~\cite{defago04}, however  the higher
price  paid  by such  algorithms  simplify  tremendously  the task  of
synchronizing persistent  data local to the replica,  specially in the
case of  failure. It also  allows for a  natural way to  aggregate the
local stable  storage of  each replica in  a global  persistent store,
without  requiring any  single replica  to assume  special  duties. In
Treplica the  \emph{ledger} abstraction of  Paxos is the  central data
structure of the whole toolkit. As a consequence, there is just a thin
software layer  between the application and  the Paxos implementation.
Thus,  Treplica doesn't  add much  overhead to  the algorithm  and our
performance data is very close to a ``pure'' Paxos implementation.

However,  there are two  factors that  characterize and  separates the
data obtained  with Treplica  from other Paxos  implementations: state
machine  execution  and operation  parallelism.   First,  we made  our
experiments using  the state machine  abstraction of Treplica,  so our
response  times  are not  equivalent  to  the  consensus latency,  but
operation execution latency. This means  that, on top of the consensus
latency,  we have to  add the  processing time  required to  apply the
command  object  to the  local  replica.   As  described in  the  next
section, we selected an application such as to minimize this cost, but
nevertheless  this latency is  present. Second,  as the  state machine
abstraction requires sequential execution  of command objects, we must
employ parallelism  internally in Treplica to avoid  the critical path
comprised  by the  Paxos ordering  and command  execution to  became a
bottleneck.  Thus,  we try  as much as  possible to pack  many command
objects  in the  same Paxos  message,  without adding  to the  overall
latency.  This way, a  multithreaded application  can obtain  a higher
throughput but the final response time deviates further from the basic
consensus latency.   As our objective  is to relatively  compare Paxos
and Fast Paxos, these effects can  be factored out as they affect both
implementations  equally.  Moreover,  both  Paxos and  Fast Paxos  are
implemented by  the same code  inside Treplica, actually a  Fast Paxos
implementation that can be configured to generate only classic rounds,
behaving  exactly like  Paxos.  Thus,  all implementation  details are
shared by the two algorithms and the comparison obtained is as fair as
possible.

\subsection{Experimental Setup}

The  experiments  were  carried  out   in  a  cluster  with  18  nodes
interconnected through the same  1Gbps Ethernet switch.  Each node has
two Intel  Xeon 2.4GHz processors, 1GB  of RAM, and a  40GB disk (7200
rpm).  System software in each node include Fedora Linux 9 and OpenJDK
Java 1.6.0 virtual machine.  We used  4 to 16 nodes in our experiments
and each node operated as a server replica and as a load generator.

The  server  replicas  run   a  simple  replicated  hash  table.   The
application  is   a  wrapper  over   the  standard  Java   hash  table
implementation,  with  the  same   API,  but  adding  replication  and
persistence support  through Treplica.  As such,  only operations that
change the internal state of  the hash table employ Treplica, the read
only operations are executed  directly.  Treplica is configured in the
server replicas to use local disk as its persistent data store, and no
network activity is expected of  each node beyond the one generated by
Treplica.

The load  generation consists in  a sequence of put  operations, where
each  operation  associates  a  sequential  integer with  a  random  5
character string. It would be  possible to interleave read with writes
in our  load, creating distinct  usage profiles.  However, due  to the
simplicity of  the application, this probably would  only increase the
observed  performance by  the proportion  of  reads used  as they  are
orders  of  magnitude  cheaper  than  writes.   Thus,  we  decided  to
concentrate on  a load composed only  of hash table  writes to analyze
the  data as  if  Treplica  were the  only  possible bottleneck.   The
generated  load is  measured in  operations per  second (op/s)  and is
generated  with  a fixed  rate  divided  equally  among all  the  load
generators of  the system. Server  replicas and load  generators share
the same hosts, but care was  taken to ensure that the load generation
wasn't  competing  with  the   application  processing  and  that  the
specified load rate was being generated.

\subsection{Experiments}

Based on the  performance expectations of Paxos and  Fast Paxos listed
in  Section~\ref{sec:expectations}  we  devised five  experiments  and
four metrics to compare both algorithms:
\begin{description}
\item[Scale up:] For  a fixed generated load of  2000 op/s we increase
  the scale  of the system from 4  to 16 replicas.  For  each point we
  count the load served in op/s and the average response time for each
  operation.
\item[Speed up:] For a fixed number of 4 and 8 replicas we increase the
  generated load from 100 op/s to  4000 op/s.  For each point we count
  the  load served  in op/s  and the  average response  time  for each
  operation.
\item[Quorum size:] We  perform the scale up and  the 8 replicas speed
  up experiments with  a modified version of Paxos  that uses a larger
  quorum than  necessary ($\lfloor  2N/3 \rfloor +  1$). We  count the
  load served in op/s.
\item[Retries  and  collisions:]  We  extract  the  number  of  failed
  consensus  instances and  collisions from  the  scale up  and the  8
  replicas  speed  up experiments.   We  count  the  number of  failed
  consensus rounds  and the number  of collisions per  total consensus
  rounds.
\item[Failures:] For a fixed number of  8 replicas and a fixed load of
  2000 op/s we simulate the  failure of a non-coordinator replica or a
  coordinator replica. We count the load served in op/s.
\end{description}

We run all experiments with  Paxos, Fast Paxos with large fast quorums
($\lceil  3N/4  \rceil$)  and  Fast  Paxos  with  small  fast  quorums
($\lfloor 2N/3 \rfloor  + 1$).  The scale up  and speed up experiments
were intended  to give  a general performance  evaluation, and  can be
used to assess if the  smaller number of communication rounds required
by Fast Paxos  and the fact that this algorithm  doesn't have a single
performance  bottleneck  make  it  more  efficient.  The  quorum  size
experiment allows us  to measure the cost of  waiting and processing a
larger  number of  messages to  achieve consensus,  indicating  if the
larger quorums required by Fast  Paxos are acceptable. The retries and
collisions  experiment  will  show  the number  of  retried  consensus
instances, an  indication of  the number of  lost messages  and timing
failures that can  be used to quantify the cost  of larger quorums and
of a single coordinator. This  experiment also shows the proportion of
collisions found  in the other  experiments to make explicit  the cost
that Fast  Paxos pays for  being optimistic.  The  failures experiment
shows how both algorithms handle  failures and if a single coordinator
can negatively affect the performance of Paxos in case of failure.

\subsection{Scale Up}

\begin{figure}[ht]
  \begin{center}
    \includegraphics[width=15cm]{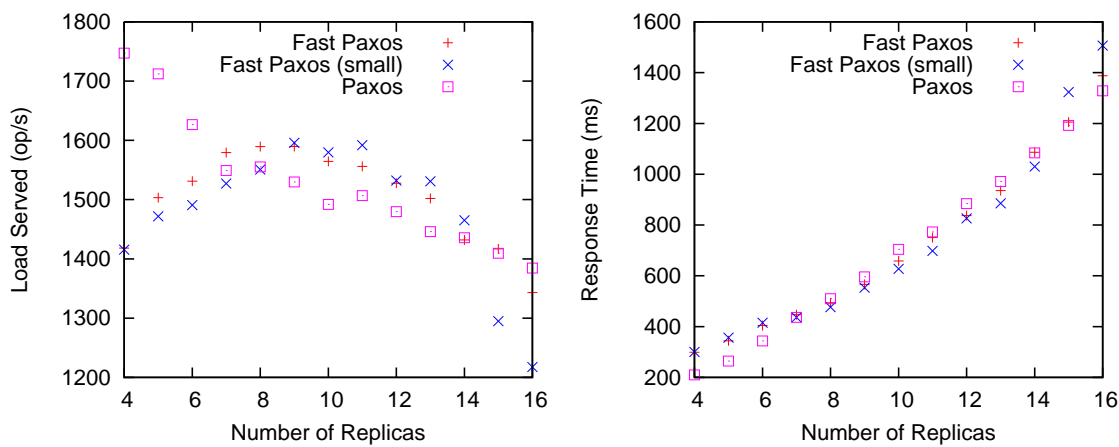}
  \end{center}
  \caption{Scale up (2000 op/s)}
  \label{fig:scaleup}
\end{figure}

Figure~\ref{fig:scaleup} shows  the data  for the scale  up experiment
with a  constant load of 2000 op/s.   The chart on the  left shows the
served operations per  second as a function of  the number of replicas
in the system. The chart on  the right shows the response time for the
same points.

The  most striking  observation  from this  experiment  is that  Paxos
outperforms Fast Paxos for small  replica numbers. Up until 7 replicas
Paxos is  better, and with more  than 7 replicas both  are roughly the
same.  Many   factors  can  justify  this  behavior,   as  pointed  in
Section~\ref{sec:expectations},  but we  believe it  is caused  by the
stabilizing  effect  the single  coordinator  creates  in the  system,
reducing timing violations.  To fully justify this supposition we need
to  analyze the  data from  the quorum  size and  retries experiments.
Another interesting behavior is the fact that Fast Paxos increases its
performance up  to a maximum at  about 9 replicas.   Again, we believe
this effect  is related to timing  violations and we  justify it using
the data  for the  retries experiments.  Both  variants of  Fast Paxos
fare similarly in all  replica configurations, with a slight advantage
for the  large quorums version.   This indicates that the  quorum size
have a role  in the performance of the algorithms but  it isn't a very
important one.   Once more, this  explanation will be verified  by the
quorum  size experiment data.   Average response  time grows  with the
number  of replicas  and all  algorithms tested  have  roughly similar
numbers.   This  is  mostly  a  consequence  of  the  fact  that  many
operations are being  ordered in the same Paxos  instance and that the
load generated is dependent on the load served.

\subsection{Speed Up}

\begin{figure}[ht]
  \begin{center}
    \includegraphics[width=15cm]{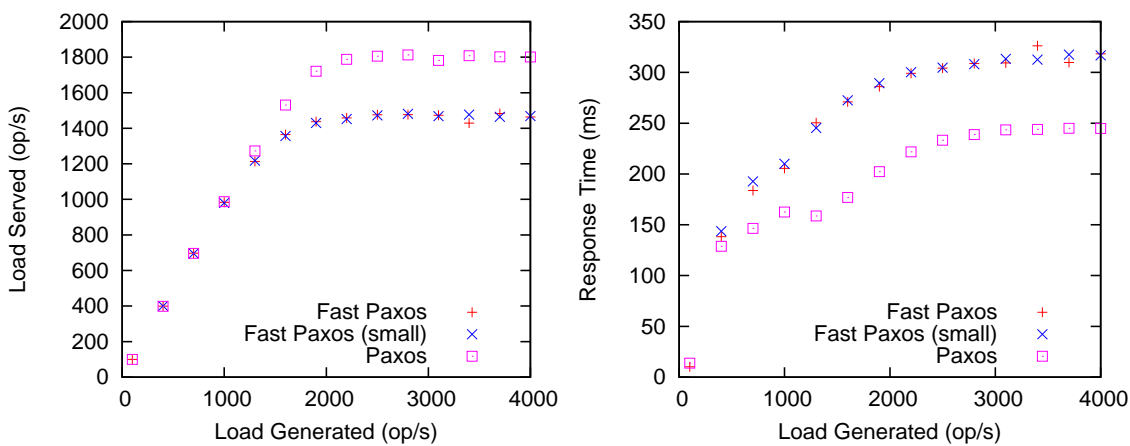}
  \end{center}
  \caption{Speedup (4 Replicas)}
  \label{fig:speedup-n04}
\end{figure}

\begin{figure}[ht]
  \begin{center}
    \includegraphics[width=15cm]{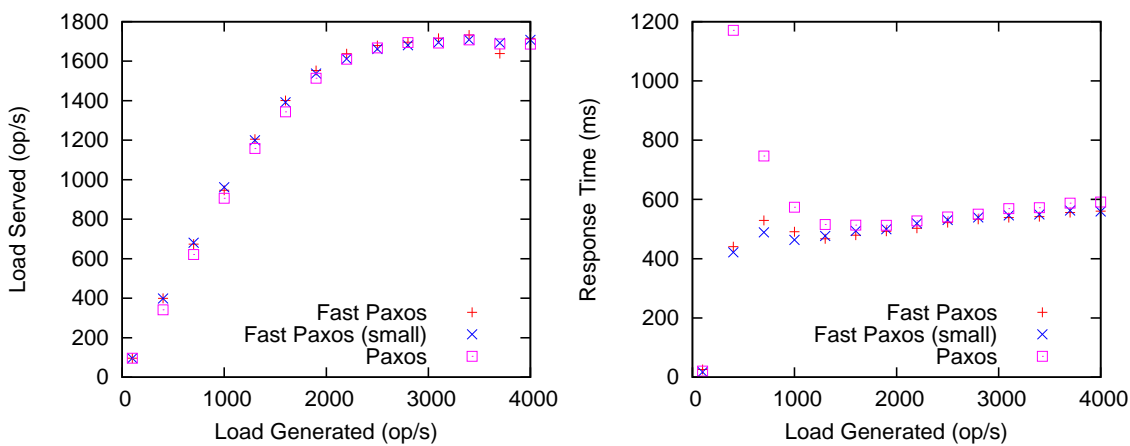}
  \end{center}
  \caption{Speedup (8 Replicas)}
  \label{fig:speedup-n08}
\end{figure}

Figure~\ref{fig:speedup-n04}   and  Figure~\ref{fig:speedup-n08}  show
data   for  the   speed  up   experiment   for  4   and  8   replicas,
respectively. In both figures, the  chart on the left shows the served
operations  per  second  as  a  function  of  the  rate  of  generated
operations per second. The chart  on the right shows the response time
for the same points.

For both 4 and 8  servers the increasing tendency of served operations
is similar.   The served load rises linearly,  following the generated
load, up  until a peak point  where it stabilizes.   This was expected
and shows  that the performance difference among  the algorithms, when
present, only shows after the  peak load is reached. Before that point
Paxos and Fast  Paxos should behave the same  way, only trading places
as  the  number  of  replica  increases  as  shown  in  the  scale  up
experiment.  The  latency charts  are more interesting.   Latency also
rises to  reach a plateau, but much  faster in the case  of 4 replicas
and even surpassing  it in the 8 replicas case.   This is explained by
the  fact that  many  operations  are bundled  in  the same  consensus
instance, and  such instances are fairly  costly.  In our  data sets a
little more than  150 consensus instances are completed  per second in
the best case. Thus, when the load is light a less aggressive bundling
takes  place  and  latency  suffers.    This  is  a  property  of  our
implementation   and  not   necessarily   will  be   found  in   other
environments.

\subsection{Quorum Sizes}

\begin{figure}[ht]
  \begin{center}
    \includegraphics[width=15cm]{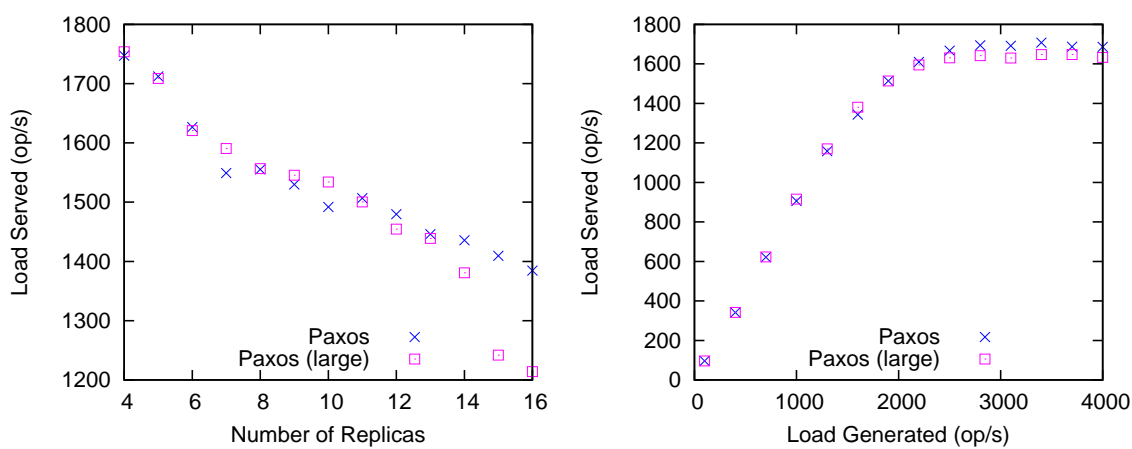}
  \end{center}
  \caption{Paxos with Large Quorums}
  \label{fig:quorums}
\end{figure}

To test the  effect of quorum sizes  we run the scale up  and 8
replicas speed up
experiments using  a modified  version of Paxos  that uses  quorums of
$\lfloor  2N/3 \rfloor  +  1$  replicas and  compare  it with  regular
Paxos.  Figure~\ref{fig:quorums} shows  the  data obtained.

Data from this experiment confirms that quorum sizes aren't a relevant
factor for performance  when the number of replicas  is moderate (less
than 15).  This  is also true for the scale up  experiment and the two
variants of Fast Paxos.  Two factors justify this finding. First, with
the  total number  of replicas  in  the 4  to 15  range, the  absolute
difference in the  cardinality of quorums is very  small, two replicas
at most.  Second, timing violations are more probable if a learner has
to receive a  message from more processes.  This  second hypothesis is
confirmed by the data  collected on consensus rounds retries presented
next.

\subsection{Retries and Collisions}

\begin{figure}[ht]
  \begin{center}
    \includegraphics[width=15cm]{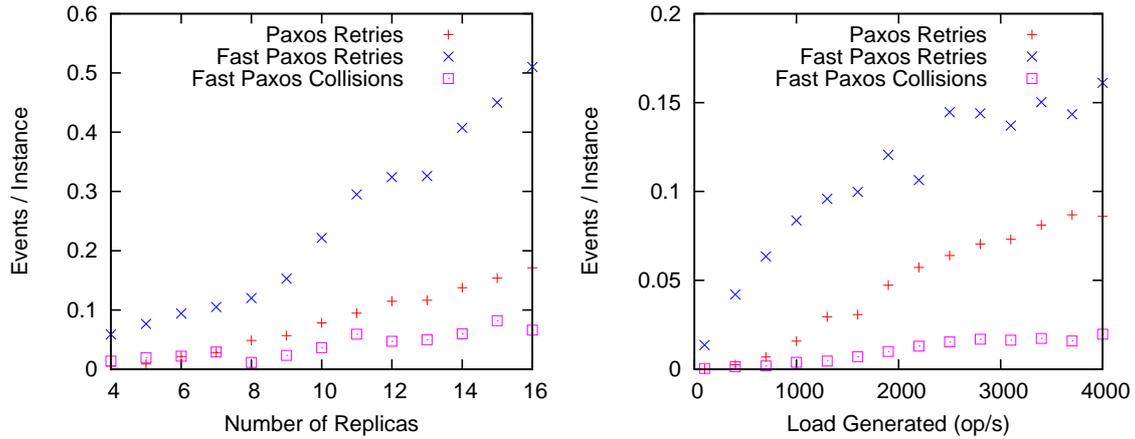}
  \end{center}
  \caption{Retries and Collisions}
  \label{fig:collisions}
\end{figure}

Figure~\ref{fig:collisions}  shows  the  number of  retried  consensus
instances for  Paxos and Fast Paxos  and the number  of collisions for
Fast Paxos  observed in  the scale up  and 8 replicas speed up  experiments. Both
numbers are  presented as  relative  values  to  the total  number  of
consensus instances executed.

This experiment  produced vital  information about the  performance of
Paxos and Fast  Paxos. The optimism of Fast  Paxos could be considered
its weak  spot and could  justify its inferior performance  with fewer
processes. However, our  data shows that collisions do  occur but they
are responsible for only a small percentage of the retried consensus instances of
Fast  Paxos. Lost  messages  or, more  likely,  timing violations  are
responsible for  the most part of consensus  failures.  Each consensus
failure triggers a regular \emph{Paxos} consensus round, even for Fast
Paxos, and this round is costly as it must execute all 2 phases of the
algorithm. The  number of failed  consensus attempts in Fast  Paxos is
sometimes  3  times larger  than  in Paxos  and  can  account for  the
decreased  performance.   The  cause  of these  timing  violations  is
probably  the fact  that timeouts  in Fast  Paxos are  managed  by all
replicas  at the  same time.   Any replica  that believes  a consensus
round should  have been finished  alerts the coordinator that  in turn
starts  a full  Paxos round,  thus we  multiply the  possibility  of a
timing violation by  the number of replicas in  the system.  In Paxos,
only the coordinator decides when a  round must be retried. It may not
be more  accurate, but the  possibility of timing failure  is smaller.
Moreover, even when a conflict does not arise in a Fast Paxos round, it
may be possible for the processes in the system to observe a ``partial
conflict'' where some, but less  than a majority, of replicas vote for
a  different operation.  In this  case, more  messages must  be timely
received for  the consensus to  be reached, increasing the  chance for
timing violations.  While this accounts for Fast Paxos limitations, it
is still necessary to explain why Paxos loses its advantage at about 8
replicas. The first cause is  that the single coordinator only acts as
a stabilizing factor as long as  it is not overloaded.  As soon as the
coordinator   gets  overloaded   it  starts   dropping   messages  and
prematurely restarting consensus rounds.

\subsection{Failures}

\begin{figure}[ht]
  \begin{center}
    \includegraphics[width=15cm]{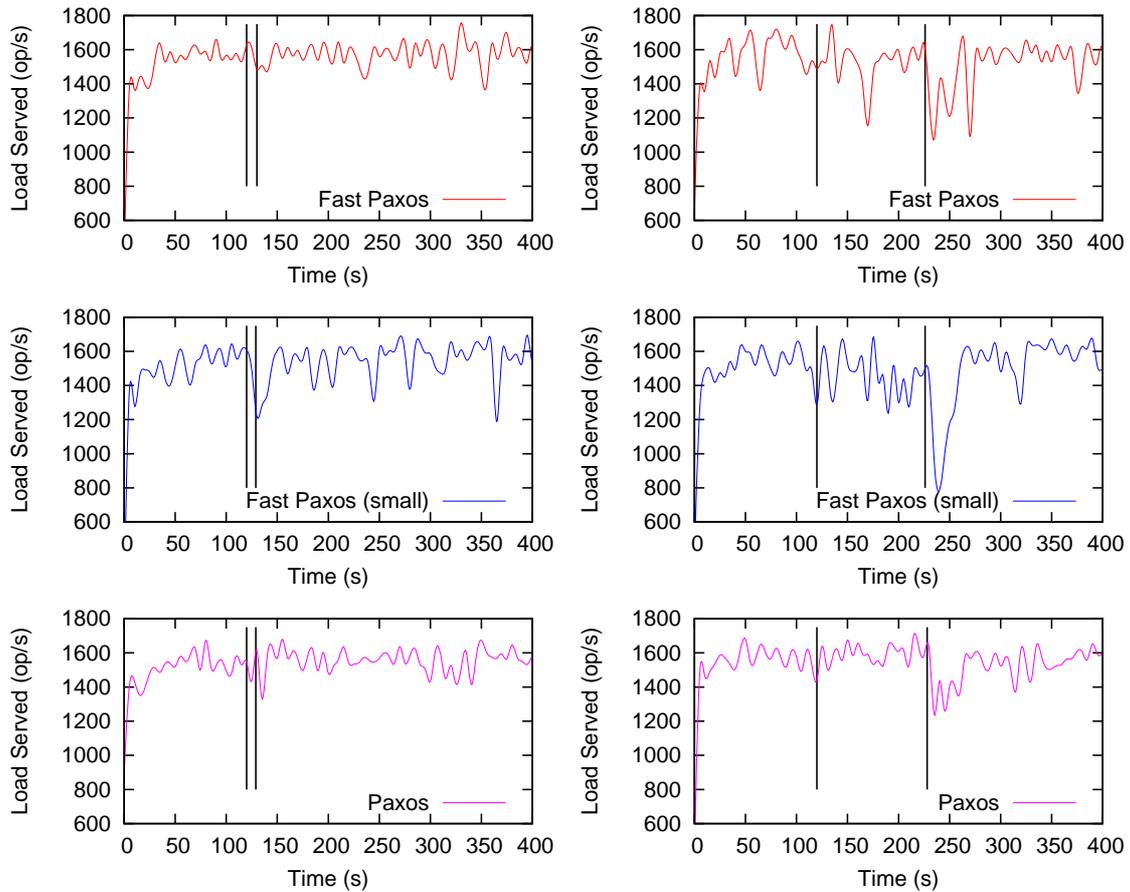}
  \end{center}
  \caption{Single Failure (8 replicas, 2000 op/s)}
  \label{fig:failure}
\end{figure}

Figure~\ref{fig:failure} shows one execution  with 8 replicas and load
of 2000 op/s that suffers the failure of a single replica. The failure
is simulated by killing the  replica at the operating system level and
by immediately  re-instantiating it back in operation.   The charts in
the left show  the failure of a regular replica and  the charts in the
right show the failure of  the coordinator replica.  In all charts the
first  vertical bar  shows the  moment  when the  replica is  forcibly
shutdown and the second bar shows the moment when the replica finishes
its  \emph{local} recovery  and starts  to coordinate  with  the other
replicas.

In both  cases it  is possible to  notice that failure  itself doesn't
impact the  throughput of the  system. This is  reasonable considering
the data from  the scale up experiment; less  replicas can potentially
give more performance.  The interesting observations is that it is the
replica reintegration  that negatively  affects the throughput  of the
system.  When a replica finishes its local recovery it has only learnt
the operations up to the moment of its failure, and must catch up with
the others  replicas. This process puts  demand on the  network and on
the coordinator as all missed  decisions are relayed to the recovering
replica. Another  intriguing aspect is  the large difference  in local
recovery  times between a  normal replica  and a  coordinator replica.
Due  to the  observed fast  local  recovery, a  normal replica  easily
reintegrates in  the system and only creates  minimal disruption.  The
failure of the  coordinator replica isn't felt any  differently by the
system,  as a  new  coordinator  is promptly  elected,  but the  local
recovery of the coordinator replica takes a longer time.  This happens
due  to the  larger  state held  in  memory by  the coordinator,  that
requires more  information to be  brought back from disk  on recovery.
As a very damaging side effect,  the longer a replica stays out of the
computation for any reason, the longer its reintegration will take and
larger  the disruption  caused by  it  will be.   Finally, all  tested
algorithms displayed a very similar behavior under failures, even when
the coordinator  has failed. This indicates that  the coordinator only
affects  the  performance of  Paxos  as  a  bottleneck in  the  steady
state. In the presence  of failures, coordinator election is performed
without interrupting the operations flow.

\section{Related Work}
\label{sec:related}

Paxos  and  Fast  Paxos  are  well understood  algorithms,  but  until
recently, seldom implemented. A  very clear and concise description of
both  algorithms can be  found in~\cite{lamport06a}.   The theoretical
performance  of  Paxos is  described  in detail  in~\cite{deprisco00}.
Probably due to  the lack of actual implementations,  one of the first
works    to    delve     in    the    Paxos    performance    employed
simulation~\cite{urban04},   and  compared   Paxos   to  Chandra-Toueg
rotating coordinator consensus algorithm~\cite{chandra96}.

Recently, motivated  by the  need of dependable  coordination services
for scalable  distributed systems, Paxos  implementations are becoming
more common and works analysing their performance are being published.
The  Chubby system  used at  Google is  described in~\cite{chandra07},
with a  some basic  performance figures. A  detailed description  of a
Paxos  implementation encompassing  all  aspects of  a complete  state
machine  replication system  can be  found in~\cite{amir08}.   In this
work it is presented a fairly complete study of the performance of the
described  implementation under  different  state machine  replication
suppositions.   A description  of  a variant  of  the Paxos  algorithm
optimized  for the  implementation  of a  distributed lock  management
system   and   an  analysis   of   its   performance   can  be   found
in~\cite{hupfeld08}.

All of  the above cited works  evaluate only Paxos.  The  only work we
have knowledge  of that attempts  to quantitatively compare  Paxos and
Fast  Paxos is~\cite{junqueira07}.   This work  employs  simulation to
study  a particular  configuration  where Fast  Paxos  doesn't have  a
better  consensus  latency  than  Paxos.  Restricted  as  the  studied
configuration  might be,  this work  showed  for the  first time  that
increased latencies of individual  messages can drastically change the
behavior of Paxos and Fast Paxos.

\section{Conclusion}
\label{sec:conclusion}

We have presented  a comparative analysis of the  performance of Paxos
and Fast  Paxos in the context  of high speed local  area networks. We
have  discovered scenarios  where Paxos  has lower  latency  than Fast
Paxos and  we showed evidence of  the cause of such  behavior.  To the
best of our knowledge this is the first such comparison.

Our experimental data  indicates that Paxos is faster  for a small set
of replicas and owns its  performance to the stability provided by its
single  coordinator.   The   Paxos  coordinator  makes  fewer  timeout
mistakes, needs to retry consensus  rounds less often and is immune to
collisions,  however  it  can  be  overloaded by  a  large  number  of
replicas.  Fast Paxos suffers  from timing failures and lost messages,
but its lack of reliance on  a single coordinator allows it to operate
more  efficiently with more  replicas.  We  have also  discovered that
quorum sizes  and collisions aren't  very determinant in  the relative
performance  of these algorithms  and that  the single  coordinator of
Paxos isn't particularly affected by failures.

As replication is used as a  device for fault tolerance, the fact that
Fast  Paxos is  more effective  with a  larger number  of  replicas is
effectively cancelled by  the fact that it requires  larger quorums of
active replicas to function. For example, a system using Paxos needs 7
replicas to tolerate  3 replica failures while Fast  Paxos requires 12
replicas to guarantee the same resilience. Thus, unless Fast Paxos can
be made more  efficient in its use of  the available network, avoiding
the timing failures observed, its use is hardly justified.

\bibliographystyle{apalike}
\bibliography{treplica-performance}

\end{document}